\documentclass{Interspeech2024}

\usepackage{multirow}
\usepackage{booktabs}
\usepackage{colortbl}
\usepackage{comment}
\usepackage{multicol}
\usepackage[normalem]{ulem}
\usepackage{hyperref}
\usepackage{url}

\interspeechcameraready 

\title{As Biased as You Measure: Methodological Pitfalls of Bias Evaluations in Speaker Verification Research} 

\name[affiliation={1,2}]{Wiebke}{Hutiri}
\name[affiliation={1}]{Tanvina}{Patel}
\name[affiliation={1}]{Aaron Yi}{Ding}
\name[affiliation={1}]{Odette}{Scharenborg}

\address{
  $^1$Delft University of Technology, The Netherlands\\
  $^2$Sony AI, Switzerland 
  }
\email{wiebke.hutiri@sony.com, t.b.patel@tudelft.nl, aaron.ding@tudelft.nl, o.e.scharenborg@tudelft.nl}


\begin{document}

\maketitle

\begin{abstract}    
Detecting and mitigating bias in speaker verification systems is important, as datasets, processing choices and algorithms can lead to performance differences that systematically favour some groups of people while disadvantaging others. Prior studies have thus measured performance differences across groups to evaluate bias. However, when comparing results across studies, it becomes apparent that they draw contradictory conclusions, hindering progress in this area. In this paper we investigate how measurement impacts the outcomes of bias evaluations. We show empirically that bias evaluations are strongly influenced by base metrics that measure performance, by the choice of ratio or difference-based bias measure, and by the aggregation of bias measures into meta-measures. Based on our findings, we recommend the use of ratio-based bias measures, in particular when the values of base metrics are small, or when base metrics with different orders of magnitude need to be compared. %Based on these results we highlight practical implications of bias, and recommend the use of ratio-based bias measures for more reliable and robust evaluation of bias in speaker recognition.
    
% Bias detection and mitigation have become an important area of research in speech processing, as datasets, processing choices and algorithms can lead to performance differences that systematically favour some groups of people while disadvantaging others. To address bias in speaker recognition, prior studies have measured performance differences across groups, and consequently evaluated bias. However, when comparing results across prior research, it becomes apparent that studies draw contradictory conclusions. This hinders our progress towards fairer speaker recognition systems. This paper investigates how different bias measures impact the outcomes of bias evaluations. We show through empirical analysis that bias evaluations are strongly influenced by the performance metric, i.e., by the choice of a ratio or difference-based bias measure, and by the aggregation of bias measures across groups. Based on these results we highlight practical implications of bias, and recommend the use of ratio-based bias measures for more reliable and robust evaluation of bias in speaker recognition.

\end{abstract}

\section{Introduction}

Speech technologies are increasingly integrated into services where reliable performance is key for human well-being and safety. One such example is speaker verification, which is used for proof-of-life verification of pensioners~\cite{morras2021bbva} and authentication of financial transactions~\cite{cox2023howvoice}. In these social-security and safety-critical applications, prediction errors can lead to technology failures that cause harms to individuals~\cite{Shelby2022SociotechnicalReduction}. In many domains that use machine learning, prediction errors have been found to be systematic, correlating with personal and demographic attributes (e.g. age, gender, accent)~\cite{Mehrabi2019Survey}. The algorithmic fairness and legal communities refer to this phenomenon as \textit{bias}. Biased speech technologies can carry significant social consequences if they assign undesirable outcomes or deny opportunities to people without reason~\cite{Friedman1996bias}. New regulations, like the EU AI Act, thus place increasing pressure on technology developers and providers to detect and mitigate bias~\cite{euaiact}. % and researchers in many application domains that use machine learning have started developing approaches for quantifying and evaluating bias~\cite{Mehrabi2019Survey}. 

Several recent studies have found evidence of bias in speaker recognition systems, for example models that are biased by speakers' gender~\cite{Hutiri2022bias, Jin2022adversarial, Fenu2020exploring, Fenu2020improving, Chen2022exploring}, nationality~\cite{Hutiri2022bias, Jin2022adversarial, Amirhossein2023ARecognition}, race~\cite{Chen2022exploring}, accent~\cite{Estevez2023StudyGroups} and age~\cite{Fenu2020improving, Fenu2020exploring}. However, despite having similar experimental setups, the studies draw varying conclusions on which groups are favoured and which are prejudiced against. For example, while one study found systems to perform worse for female speakers and non-US nationals~\cite{Hutiri2022bias}, another study that trained and evaluated on the same dataset found the opposite; that models perform better for females, and better for UK nationals than for US nationals~\cite{Jin2022adversarial}. %Yet another study claimed that models for languages other than English tend to perform better for males than females when the age range of speakers is the same~\cite{Fenu2020exploring}. 
One reason for these divergent claims is that studies use different metrics and measures to compare performance errors across groups of people.
% approaches for evaluating bias. On the one hand, bias evaluations in speaker verification research are frequently motivated by constraints, such as the metadata available for comparing performance across demographic attributes, or a preference for binary groups (e.g. male/female) to suit the input dimensions of established bias measures. On the other hand, the metrics and measures that are chosen to compare performance errors impact bias evaluations in their own right.

In this paper we study how the metrics and measures used to quantify bias impact the validity of bias evaluations of speaker verification systems. First, we introduce terminology to distinguish base metrics from bias measures and meta-measures. We then compare three bias measures and two meta-measures from the literature, showing empirically how they lead to different bias evaluation outcomes. Finally, we demonstrate with a thought experiment how biased speaker verification systems can impact people in a real-world application, and why reliable bias evaluations are important to prevent this. Based on these insights we make recommendations for more reliable bias evaluations that can lead to fairer and more inclusive speaker verification systems. 

% In speech technology, studies on bias however are only nascent and lack methodological consistency. In speaker recognition research, several studies have highlighted sources of bias. For example, model performance suffers for underrepresented groups in training datasets~\cite{Shen2022improving, Fenu2021Fair, Estevez2023StudyGroups}, and groups with small amounts of speakers in evaluation datasets show high performance variance~\cite{Hutiri2022bias}. Despite progress in the area of bias evaluation in the speaker recognition domain, studies that investigate bias suffer from methodological variability. For example, several studies have examined bias due to nationality~\cite{Jin2022adversarial, Hutiri2022bias}, age~\cite{Fenu2021Fair, Fenu2020exploring}, language~\cite{Fenu2020exploring} and race~\cite{Chen2022exploring}. However, group categories vary across studies and are frequently motivated by constraints such as available metadata or the input requirements for the bias measures, irrespective of whether this categorisation leads to meaningful insights about bias. In this paper we study the impact of bias evaluation methodologies, in particular the metrics and measures used to quantify bias, on bias evaluation outcomes.
\section{Background and Related Work}

% \subsection{Bias and Fairness}

% \textit{Bias} in its simplest form refers to a skewed or slanted perspective. Biased technologies can carry significant social consequences if they systematically assign undesirable outcomes or deny opportunity to people without reason~\cite{Friedman1996bias}. In line with this framing, bias in machine learning (ML) frequently implies that a model or system produces unequal error rates for individuals or groups, based on their personal attributes (e.g. age, language, accent, health, country of origin). 
% \textit{Fairness} is often considered as the aspirational antithesis to \textit{unfairness}, where unfairness can be caused by biases in the ML development process, such as unrepresentative training data, optimisation objectives or data labelling choices~\cite{Pessach2023review}. \textit{Discrimination} concerns the effects of the outcome of a decision-making process~\cite{mittelstadt2016ethics}. In many countries anti-discrimination, or non-discrimination, is a legal requirement and decision-making processes need to treat individuals and groups of people equally, taking protected personal attributes into account~\cite{wachter2021bias}. In contrast to its legal meaning and usage in algorithmic fairness, in speaker recognition discrimination refers to the ability to tell different speakers apart, and is a desirable property of models~\cite{Brummer2006application}. 

Base metrics, bias measures and meta-measures are essential components of bias evaluations~\cite{hutiri2023designai}. \textit{Bias measures} quantify and thus measure bias for the purpose of bias detection (or diagnosis) and mitigation (or intervention)~\cite{barocas2019fairness}. This paper focuses on the former. When used for detection purposes, bias measures can be applied during model development or post-hoc to test models and applications in order to gain insights into the limits of their performance. Most bias measures are calculated from statistical \textit{base metrics} that quantify model performance or prediction error rates. Common base metrics used in speaker recognition are the false positive (FPR) and false negative (FNR) error rates, equal error rate (EER) and the minimum detection cost (minCDet)~\cite{greenberg2020two}. Base metrics can be disaggregated across groups of people to evaluate model performance across demographic or other protected attributes~\cite{verma2018fairness}. To compare model performance across groups, bias measures calculate ratios or differences between the base metrics of a group and a reference group, or overall performance. \textit{Meta-measures} aggregate bias measures across groups into a single score for a model~\cite{Lum2022} to support the comparison of bias across different models. Like bias measures, meta-measures can be computed for different base metrics. 

% \subsection{Measuring Bias in Speaker Verification}

\begin{table*}[!htb]
\footnotesize
\setlength{\tabcolsep}{3pt}
\caption{Bias measures evaluated in this study}
\label{tab:bias_measures}
    \centering
    \begin{tabular}{p{0.12\linewidth}p{0.29\linewidth}p{0.32\linewidth}cp{0.06\linewidth}}
    \textbf{Name} & \textbf{Description} & \textbf{Equation} & \textbf{Reference} & \textbf{In meta-measure}\\ \midrule
    Group-to-min Difference & Distance between the base metric (b) of a group (g) and the base metric of the best performing group (m) & $\mathnormal{G2min\ diff(b)_g = b_{g} - b_{m} }$ & \cite{Shen2022improving, Jin2022adversarial, Fenu2020exploring} & FDR \\ %$\mathnormal{distance\ to\ min(b)_g = b_{g} - b_{m} }$
    Group-to-average Ratio & Ratio between a group's base metric and the average base metric value across all groups & $\mathnormal{G2avg\ ratio(b)_g =\frac{b_{g}}{b_{average}}}$ & \cite{Hutiri2022bias} & - \\
    Group-to-average log Ratio & Negative log of the Group-to-average Ratio & $\mathnormal{G2avg\ log\ ratio(b)_g =-ln\left(G2avg\ ratio(b)_g\right)}$ & \cite{Hutiri2023TinyWorkflows} & NRB \\
    \end{tabular}
\vspace{-2em}
\end{table*}

% Several recent studies have found evidence of bias in speaker recognition systems, for example models that are biased by speakers' gender~\cite{Hutiri2022bias, Jin2022adversarial, Fenu2020exploring, Fenu2020improving, Chen2022exploring}, nationality~\cite{Hutiri2022bias, Jin2022adversarial, Amirhossein2023ARecognition}, race~\cite{Chen2022exploring}, accent~\cite{Estevez2023StudyGroups} and age~\cite{Fenu2020improving, Fenu2020exploring}. Despite these studies having similar experimental setups, their conclusions of which groups are favoured and which are prejudiced against, vary. For example, while one study found systems to be biased against female speakers and non-US nationals~\cite{Hutiri2022bias}, another study found that models perform better for females than males and better for UK nationals than for US nationals~\cite{Jin2022adversarial}. Yet another study claimed that models for languages other than English tend to perform better for males than females when the age range of speakers is the same~\cite{Fenu2020exploring}. 

The most common base metric used to measure performance in studies that investigate bias in speaker recognition is the EER~\cite{Shen2022improving, Jin2022adversarial, Fenu2020exploring, Fenu2021Fair, Peri2022train}. Other base metrics that have been considered are the minCDet~\cite{Hutiri2022bias}, the log likelihood-ratio cost function (Cllr)~\cite{Estevez2023StudyGroups}, the FNR at a FPR of 1\%~\cite{Fenu2021Fair} the false accept rate (same as FPR) and the false reject rate (same as FNR)~\cite{Peri2022train}. Bias measures can be classified broadly as difference-based and ratio-based measures.  Most studies use difference-based measures calculated either from the EER~\cite{Shen2022improving, Jin2022adversarial, Fenu2020exploring} or from statistical fairness measures in ML~\cite{Fenu2021Fair, Amirhossein2023ARecognition}. However, with the exception of the equalized odds ratio, most statistical fairness measures consider performance disparities either due to false positive or due to false negative errors. As speaker verification systems trade off the FPR and FNR, this limits the utility of statistical fairness measures for bias evaluations in the speaker verification domain. Only one study used a ratio-based bias measure with the EER and minCDet base metrics~\cite{Hutiri2022bias}. Studies that use a meta-measure have adopted the Fairness Discrepancy Rate (FDR)~\cite{Estevez2023StudyGroups, Peri2022train}, which was first proposed to assess fairness in biometric verification systems~\cite{DeFreitasPereira2022fairness}. %Given these diverse approaches to measuring bias, we now set out to study the effect of base metrics, bias measures and meta-measures on the outcomes of bias evaluations of speaker verification systems.
\section{Method}

This section defines the bias and meta-measures that we compare, and describes the experimental setup. The software used for the analysis has been released as a package on PyPI\footnote{\url{https://pypi.org/project/bt4vt/}}.

\subsection{Bias and Meta-measures}
Table~\ref{tab:bias_measures} defines three bias measures from the literature that we compare in this study: the \textit{Group-to-min (G2min) Difference}, the \textit{Group-to-average (G2avg) Ratio} and the \textit{G2avg log Ratio}. They can be used with any base metric. In addition we compare two meta-measures, the Fairness Discrepancy Rate (FDR) and the Normalised Reliability Bias (NRB), which we define below. 

The FDR~\cite{DeFreitasPereira2022fairness} performs a pairwise comparison of the FPR and FNR differences across groups at a threshold $\tau$. For each error rate it selects the pair with the maximum difference (i.e. the maximum value of the \textit{G2min Difference} bias measure). The maximum differences are then weighted by $\alpha$, and combined into a joint measure, the FDR.

\begin{equation}
\footnotesize
\label{eq:fdr}
\begin{split}
    max_{\Delta FPR}(\tau) &= max\left(G2min\ diff(FPR(\tau))_G\right)\\
    max_{\Delta FNR}(\tau) &= max\left(G2min\ diff(FNR(\tau))_G\right) \\ 
    FDR(\tau) &= 1 - ( \alpha \times max_{\Delta FPR}(\tau) \\
     & + (1 - \alpha) \times max_{\Delta FNR}(\tau) ) ; 0 <= \alpha <= 1
\end{split}
\end{equation}

\noindent The FDR ranges from 0 (most biased) to 1 (least biased). It can be evaluated at different thresholds $\tau$, which produce different design error rates $FPR_{avg}$. Choosing $\tau$ is a form of choosing a base metric, as each threshold produces a unique $(FPR_{avg},\ FNR_{avg})$ pair. The FPRs and FNRs of groups will deviate from those of the system average, unless the system is unbiased. The system can further be evaluated for different weights $\alpha$. When $\alpha = 0$ the FDR only accounts for FN errors. When $\alpha = 1$, only FP errors are evaluated. 

As a second meta-measure we consider Reliability Bias, which was proposed to measure quality-of-service harms in on-device keyword spotting systems~\cite{Hutiri2023TinyWorkflows}. The measure calculates the sum of the absolute values of the \textit{G2avg log Ratio}. To make the Reliability Bias meta-measure comparable across variable numbers of groups, we normalise it by dividing by the number of groups ($G$). The Normalised Reliability Bias (NRB) in Equation 2 has a lower bound of 0 when the performance across all groups is equal and the model is unbiased. The upper limit is infinite. The higher the score, the greater the difference between group and average performance, and the more biased the model. Note that this interpretation is opposite to that of the FDR. 

\begin{equation}
\label{eq:nrb}
    \mathnormal{NRB(b) = 1/G \sum_{g=1}^{G} \lvert G2avg\ log\ ratio(b)_{g} \rvert}
\end{equation}

\subsection{Experiment Setup}
To investigate the bias and meta-measures, we use a pre-trained end-to-end ResNet-34 speaker verification model from the Clova baseline\footnote{We use the ``performance-optimized" model.}~\cite{heo2020clova}, trained on the VoxCeleb2 dataset~\cite{Nagrani2020voxceleb} as a black-box predictor. The model is evaluated on two evaluation sets constructed from trial pairs in the VoxCeleb1 dataset: VoxCeleb1-H and VoxCeleb1-I. VoxCeleb1-H consists exclusively of trial pairs where speakers have the same nationality and gender. However, prior research showed that across nationalities 8\% - 17\% of same speaker pairs in VoxCeleb1-H use trial pairs that come from the same voice recording, making these comparisons trivial~\cite{Hutiri2022design}. Moreover, the proportion of trivial same speaker pairs is not the same across nationalities, which results in a skewed evaluation setup. We thus also evaluate on the VoxCeleb1-I trial pairs proposed in~\cite{Hutiri2022design}. %The set contains 520 same-speaker and 520 different-speaker utterance pairs for each speaker. Same-speaker pairs come from different recordings, while different-speaker pairs come from speakers with the same gender and nationality. This setup ensures that there is an equal number of evaluation pairs for each speaker and that the trial pairs are of equivalent difficulty across speakers.

We limit our bias evaluation to groups that can be constructed from demographic metadata released with VoxCeleb1, namely binary \textit{gender} (male, female) and the intersection of \textit{gender and nationality}, for the following nationalities: Ireland, India (IN), USA (US), Australia (AUS), Canada, UK, Norway (NO) and Germany (DE). 
\section{Results}

We now present our results, starting with an overview of model performance and disaggregated base metrics across groups. Next, we anaylse how the base metrics and bias measures, and then the meta-measures impact the outcomes of the bias evaluation. Our analysis is available as a jupyter notebook\footnote{\url{https://github.com/wiebket/measuring_bias_speech/}}.

The average EER and minCDet values of the speaker verification model are (2.402, 0.008) and (3.657, 0.012) for VoxCeleb1-H and -I respectively. While the performance measures are 50\% greater (i.e. the model performs worse) when evaluating on the more challenging conditions of the VoxCeleb1-I set, the overarching trends are similar. We present our analysis on VoxCeleb1-I going forward. Table~\ref{tab:sv_results} shows disaggregated base metrics for \textit{gender} groups in the left column labelled `All', and for intersectional \textit{gender + nationality} groups in the remaining columns. Due to space constraints we only show results for the best and worst performing nationalities (IN, US, AUS, NO, DE). 

\begin{table}[!hbt]
\footnotesize
\setlength{\tabcolsep}{5pt}
\centering
\caption{Disaggregated EER and minCDet base metrics on VoxCeleb1-I for \textbf{gender} (col. `All') and \textbf{gender + nationality} groups (\textbf{bold} is best performing base metric in group).}
\label{tab:sv_results}
% \resizebox{\linewidth}{!}{
\begin{tabular}{p{0.18\linewidth}|c|ccccc}
\toprule
\multicolumn{7}{c}{\textbf{Male}} \\ \midrule
\textbf{Base metric} & \textbf{All} & \textbf{IN} & \textbf{US} & \textbf{AUS} & \textbf{NO}& \textbf{DE} \\ \midrule
\textbf{EER} & \textbf{3.581} & 3.218 & 2.999 & 4.362 & 8.210 & 3.013 \\ 
\textbf{minCDet} & \textbf{0.011} & 0.018 & 0.010 & 0.012 & 0.025 & \textbf{0.009} \\ \midrule
\multicolumn{7}{c}{\textbf{Female}} \\ \midrule
\textbf{EER} & 3.757 & 7.028 & 3.250 & \textbf{2.788} & 4.588 & 10.641 \\
\textbf{minCDet} & 0.012 & 0.023 & 0.011 & 0.011 & 0.014 & 0.019 \\
\hline
\end{tabular}
% }
\vspace{-2em}
\end{table}

\noindent For \textit{gender} groups, the model performs better for males than females for both base metrics. For \textit{gender + nationality} groups, the EER is lowest for Australian females, and the minCDet lowest for German males. For male and female \textit{gender + nationality} groups there are groups with substantially worse than average performance. For example, Norwegian males have an EER that is 2.7 times that of US males. Similarly, German females have 3.8 times the EER of Australian females, but only 1.7 times the minCDet. These results show that the performance of the model varies significantly across genders and nationalities, implying that it is biased. However, the results also suggest that \textbf{the extent of bias depends on the base metric}. 

%Table~\ref{tab:sv_results} shows the results per speaker group. The rules used to generate the VoxCeleb1-H and VoxCeleb1-I datasets precluded the creation of a German male and Irish female speaker group respectively, due to insufficient data for these groups. The minCDet base metric for groups was calculated at the threshold which produces the average minCDet value.

% \begin{table}[hbt]
% \centering
% \caption{\textbf{Average} speaker verification system performance for EER and minCDet base metrics.}
% \label{tab:sv_avg_performance}
% \begin{tabular}{c|cc}
% \toprule
%    \textbf{Base metric} & \textbf{VoxCeleb1-H} & \textbf{VoxCeleb1-I} \\ \midrule
%    \textbf{EER} & 2.402 & 3.657 \\
%    \textbf{minCDet} & 0.008 & 0.012 \\
% \bottomrule
% \end{tabular}
% \end{table}

\subsection{Impact of Base Metrics and Bias Measures}
\label{ss:impact_bms}
Table~\ref{tab:sv_biasmeasures} shows the bias measures evaluated for the EER and minCDet base metrics across \textit{gender} and best and worst performing \textit{gender + nationality} groups. For \textit{gender} groups (i.e. column `All') the model shows preference for the male group across all base metrics and bias measures. The ratio-based bias measures for males have a \textit{G2avg Ratio} less than 1 and a positive \textit{G2avg log Ratio}, indicating that performance for this group is always better than average. For the female group the inverse is true: the \textit{G2avg Ratio} is greater than 1 and the \textit{G2avg log Ratio} is negative, indicating that performance is always worse than average. The difference-based \textit{G2min Difference} evaluates to 0 for the male group, which has the smaller error rates and is thus used as reference. For the female group, the EER and minCDet base metrics are two orders of magnitude apart. This makes it difficult to compare the \textit{G2min Difference} across the base metrics to establish their impact on the measure. The two ratio-based bias measures, by contrast, are invariant to the order of magnitude of the base metric. We can thus compare the bias measures for the EER and minCDet base metrics to confirm what we observed in Table~\ref{tab:sv_results}, namely that the extent of bias depends on the base metric used to measure performance. 

%For the \textit{gender + nationality} group one best base metric was selected from male and female groups to calculate the \textit{Group-to-min Difference}. Similarly, the average base metric was calculated from all male and female groups for the Group-to-average ratios.

\begin{table}[htb]
\footnotesize
\setlength{\tabcolsep}{3.8pt}
\caption{Bias measures for \textbf{gender} (`All'), best and worst performing \textbf{gender + nationality} groups (\textbf{bold} is most favoured).}
\label{tab:sv_biasmeasures}
\begin{tabular}{p{0.18\linewidth}|p{0.13\linewidth}|c|cccc}
\toprule
\multicolumn{7}{c}{\textbf{Male}} \\ \midrule
 \textbf{Bias meas.} &  \textbf{Base} & \textbf{All} & \textbf{IN} & \textbf{US} & \textbf{AUS} & \textbf{DE} \\ \midrule
 G2min Diff. &  EER & 0.000 & 0.429 & 0.211 & 1.573 & 0.224 \\
  &  minCDet & 0.000 & 0.010 & 0.001 & 0.003 & \textbf{0.000} \\ \hline
G2avg Ratio & EER & 0.979 & 0.880 & 0.820 & 1.193 &  0.824 \\
 &  minCDet & 0.954 & 1.571 & 0.863 & 1.046 &  \textbf{0.749} \\ \hline
G2avg &  EER & 0.021 & 0.128 & 0.198 & -0.176 & 0.194 \\
 log Ratio &  minCDet & 0.047 & -0.452 & 0.148 & -0.045 &  \textbf{0.289} \\ \midrule
\multicolumn{7}{c}{\textbf{Female}} \\ \midrule
 % \textbf{Bias measure} &  \textbf{Base} & \textbf{All} & \textbf{IND} & \textbf{US} & \textbf{AUS} & \textbf{DE} \\ \hline
G2min Diff. &  EER & 0.176 & 4.240 & 0.462 & \textbf{0.000} & 7.853 \\
 &  minCDet & 0.001 & 0.015 & 0.002 & 0.002 & 0.011 \\ \hline
G2avg Ratio &  EER & 1.027 & 1.922 & 0.889 & \textbf{0.762} & 2.909 \\
 &  minCDet & 1.059 & 1.986 & 0.937 & 0.945 & 1.662 \\ \hline
G2avg &  EER & -0.027 & -0.653 & 0.118 & \textbf{0.271} & -1.068 \\
 log Ratio &  minCDet & -0.057 & -0.686 & 0.065 & 0.056 & -0.508 \\ \hline
\end{tabular}
\vspace{-2em}
\end{table}

The impact of the bias measures on bias evaluations becomes more evident for the multicategory \textit{gender + nationality} groups (cols IN, US, AUS, DE). Firstly, we observe that \textbf{bias measures preserve the ranking of groups by performance for a particular base metric, but can change it across base metrics}. For example, Australian females have the lowest EER and are used as reference for the \textit{G2min Difference}. This group also has the lowest \textit{G2avg Ratio} of 0.762 and the highest \textit{G2avg log Ratio} of 0.271. All bias measures thus show that this group is strongly favoured when using the EER base metric. However, when computing bias measures with the minCDet, the reference for the \textit{G2min Difference} changes to German males, who have the lowest minCDet value. When analysing the \textit{G2avg Ratio} and \textit{G2avg log Ratio}, German males are now the most favoured group, followed by US males, US females and only then Australian females. In addition to changing the order of preference, \textbf{a change in base metric can also lead to a different conclusion about bias}, as is the case with Indian males who change from being favoured to being prejudiced against when the base metric changes from the EER to the minCDet. %These results correspond with the disaggregated base metrics in Table~\ref{tab:sv_results}.

\subsection{Impact of Meta-measures}
\begin{figure}[!b]
\vspace{-1em}
    \centering
    \includegraphics[width=\linewidth]{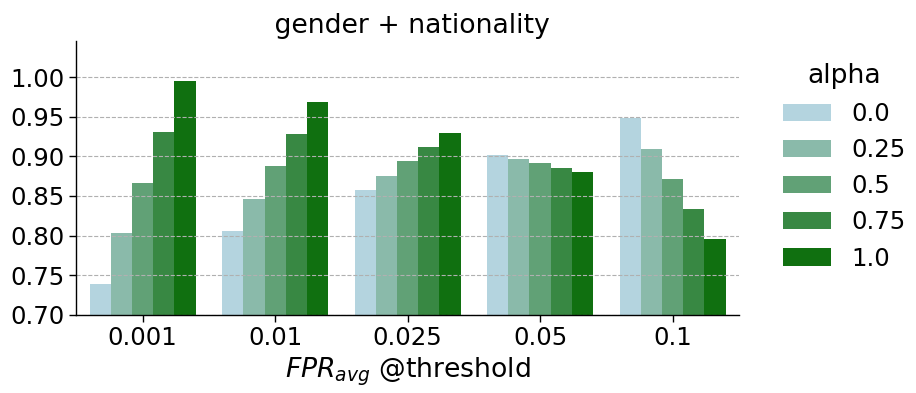}
    \vspace{-1.5em}
    \caption{\textbf{FDR meta-measure} for \textbf{gender + nationality} groups. The FDR is calculated for different $\alpha$ and for systems calibrated to thresholds that produce pre-determined $FPR_{avg}$. $\alpha=0$ only considers the \textit{FNR}, while $\alpha=1$ only considers the \textit{FPR}.}
\label{fig:fdr_at_fpr+alpha}
\vspace{-1em}
\end{figure}

We now compare the FDR and NRB meta-measures to consider how aggregating bias measures into a single meta-measure further impacts bias evaluations. We show results for \textit{gender + nationality} groups. \textit{Gender} groups, which are not shown, follow a similar but weaker trend. Figure~\ref{fig:fdr_at_fpr+alpha} visualises the results of a bias evaluation using the FDR from Equation 1 with different $\alpha$ and $\tau$. We selected $\tau$ that calibrate the system to $FPR_{avg} = \{0.001, 0.01, 0.025, 0.05, 0.1\}$, and evaluated the FDR at $\alpha=\{0, 0.25, 0.5, 0.75, 1\}$. The figure shows that at a small $FPR_{avg}$ (e.g. 0.001) the FDR approaches 1 (i.e. least bias) as $\alpha$, which increases the weight of the FPR, increases. This implies that the FNR determines the FDR bias value at small $FPR_{avg}$. This trend is reversed for systems calibrated to larger $FPR_{avg}$ (e.g. 0.1), where larger $\alpha$ reduce the FDR, implying that the FPR determines bias. 

% This means that increasing the weight of the FPR base metric reduces bias, or conversely that the FNR increases bias.  

 %At $FPR_{avg}=0.1$ the opposite is the case: increasing $\alpha$ from 0 (light blue) to 1 (dark green) reduces the FDR, which implies that the FPR increases the system's bias. This effect is exacerbated for the \textit{gender + nationality} groups, where it is clear that for systems calibrated to \textbf{smaller $FPR_{avg}$}, e.g. 0.001, smaller $\alpha$, which is to say the \textbf{FNR, increase bias}. 

% For all thresholds and $alpha$ the FDR is very close to 1 when considering \textit{gender} groups. This implies that there is almost no difference in performance for males and females. On close examination it is possible to see

\begin{figure}[t]
    \centering
    \includegraphics[width=0.95\linewidth]{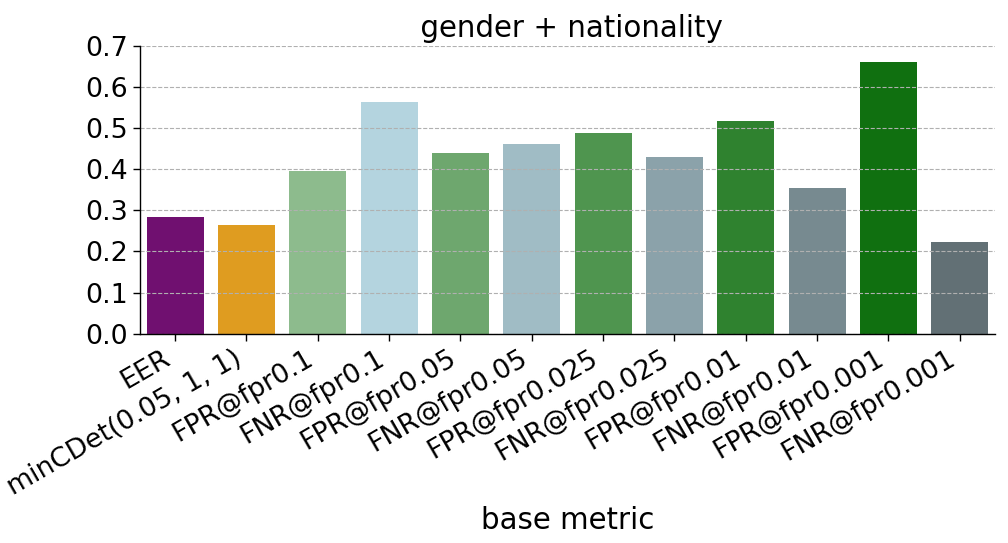}
    \vspace{-0.5em}
    \caption{\textbf{NRB meta-measure} for \textbf{gender + nationality} groups. The meta-measure is calculated for different base metrics.}
\label{fig:reliability_bias_at_metrics}
\vspace{-1.5em}
\end{figure}

Next, we conduct a similar evaluation for the NRB from Equation 2, testing it with a range of base metrics; the EER, minCDet, and FPRs and FNRs at systems calibrated to $FPR_{avg}=\{0.001, 0.01, 0.025, 0.05, 0.1\}$. These $FPR_{avg}$ values have been chosen to correspond with those evaluated for the FDR. Figure~\ref{fig:reliability_bias_at_metrics} shows the NRB values for each base metric. Moving from left to right, $FPR_{avg}$ decreases and the NRB increases (i.e. shows greater bias) for the FPRs (green). For the FNRs (blue-grey) the opposite is true: the NRB decreases as $FPR_{avg}$ decreases. For example, at $FPR_{avg}=0.1$ (i.e. left side of the chart), the NRB is lower when calculated with the FPR than with the FNR. However, at $FPR_{avg}=0.001$ (i.e. right side of the chart), the NRB is substantially greater when calculated with the FPR than the FNR.  %Calibration to a smaller $FPR_{avg}$ (e.g. 0.001) thus leads to the FPR increasing the bias value, while calibration to a larger $FPR_{avg}$ (e.g. 0.1), leads to the FNR increasing bias. 

\textbf{The bias evaluations with the FDR and NRB thus lead to contradictory conclusions.} To illustrate this, consider a hypothetical system that will be used in an application that requires high security. This necessitates a low FPR, and the system is thus calibrated to $FPR_{avg}=0.001$. Bias is then evaluated specifically for the FPR base metric, given its importance to the use case. We obtain the FDR at $\alpha=1$, which weights the meta-measure to only consider bias due to the FPR. From Figure~\ref{fig:fdr_at_fpr+alpha} we estimate a bias value of $\sim$0.99, which suggests that the model contains minimal bias and is safe to use. Next, from Figure~\ref{fig:reliability_bias_at_metrics} we estimate the NRB of this system as $\sim$0.65 (green bar on the right). This value indicates that substantial performance discrepancies exist across speakers with different genders and nationalities. The system should not be used, as the security of some groups will be severely jeopardised. 

\subsection{Which meta-measure is correct?}

% say something about impact of evaluation datasets on bias evaluations, group design, only having studied one pretrained model here 

How can these two meta-measures lead to opposite conclusions about bias? To investigate this, we decompose the meta-measures into their constituent bias measures and base metric in Table~\ref{tab:fpr0.001_basemetric}. %shows the FPR base metric, and the \textit{G2min Difference} and \textit{G2avg log Ratio} bias measures at a design $FPR_{avg}=0.001$. The results are disaggregated across \textit{gender + nationality} groups. 
Given the small FPR values, the \textit{G2min Difference} values, which contribute to the calculation of the FDR, are equally small and therefore insensitive to small performance differences across groups. This observation is affirmed by Equation~\ref{eq:fdr}, which shows that the FDR is primarily influenced by the order of magnitude of the base metric, and secondly by the value of $alpha$. When the base metric is small, the FDR is thus prone to underestimate performance differences across groups. By contrast, the \textit{G2avg log Ratio}, which is used to calculate the NRB, captures a relative relationship and is unaffected by the order of magnitude of the base metric. The high bias value of the NRB thus reflects the disparity in FPRs across groups shown in Table~\ref{tab:fpr0.001_basemetric} and correctly identifies the system as biased. 

\begin{table}[t]
\footnotesize
\setlength{\tabcolsep}{3pt}
\centering
\caption{Disaggregated FPR, G2min Difference and G2avg log Ratio at design $FPR_{avg}=0.001$ (\textbf{bold} $>$ 0.001).}
\label{tab:fpr0.001_basemetric}
\begin{tabular}{c|cccc}
\toprule
\multicolumn{5}{c}{\textbf{Male}} \\ \midrule
\textbf{Base metric / Bias measure} & \textbf{IN} & \textbf{US} & \textbf{AUS} & \textbf{DE} \\ \midrule
\textbf{FPR@fpr0.001} & \textbf{0.005} & 0.000 & 0.001 & \textbf{0.002} \\ \cmidrule{2-5}
\textbf{G2min Difference} & 0.005 & 0.000 & 0.001 & 0.002 \\ 
\textbf{G2avg log Ratio} & 1.659 & -0.912 & 0.000 & 0.654 \\ \midrule
\multicolumn{5}{c}{\textbf{Female}} \\ \midrule
\textbf{FPR@fpr0.001} & \textbf{0.003} & 0.001 & \textbf{0.002} & \textbf{0.001} \\ \cmidrule{2-5}
\textbf{G2min Diff} & 0.003 & 0.000 & 0.001 & 0.001 \\
\textbf{G2avg log Ratio} & 1.201 & -0.475 & 0.472 & 0.249 \\
\bottomrule
\end{tabular}
\vspace{-1em}
\end{table}

While we have demonstrated that the NRB correctly identifies bias, it remains important to assess if the seemingly small differences in FPR that we observe in Table~\ref{tab:fpr0.001_basemetric} matter. To explore this, we consider a scenario where an attacker gains access to a device with the previously described speaker verification system. They attempt to access sensitive information on the device by invoking the speaker verification system once a minute (i.e. 60 times / hour). At the design FPR of 0.001 they have a 1 in a 1000 chance of gaining access to the system. After 17 hours (i.e. 1020 attempts), they are likely to succeed. If the device belonged to an Indian male, the FPR of 0.005 would now grant the attackers a 1 in 200 chance of success. This means that they only need to attack the system for 3.5 hours to gain access. This increased exposure to successful attacks presents greater risk of harm to groups that have worse than average performance. The NRB thus correctly identifies this system as biased, while the FDR misrepresents the potential risk.
\section{Discussion and Limitations}
% Bias becomes consequential when disparate error rates have real-world impacts that disadvantage some people more than others.
The results in this paper demonstrate that bias evaluations are important to prevent unfair speaker verification systems. However, they also show that evaluations are highly influenced by the choice of base metrics, bias measures and meta-measures. We highlight that the performance ranking of groups depends on the base metric used to measure performance. We further show that bias measures, which are calculated from base metrics, are affected by the order of magnitude of base metrics. Importantly, difference-based measures such as the \textit{G2min Difference} cannot be compared across base metrics with different orders of magnitude, and lack sensitivity when base metrics are small. These shortcomings affect meta-measures based on difference-based bias measures, such as the FDR. We thus recommend the use of ratio-based measures, which are invariant to the magnitude of the base metric, and meta-measures like the NRB that are calculated from ratio-based bias measures. We find the \textit{G2avg log Ratio}, which is centered around 0, easier to compare across multiple groups than the \textit{G2avg Ratio}. 

The base metrics and bias measures that we investigate in our empirical study are those that are most frequently used in speaker verification bias evaluations. However, they are not the only metrics and measures that can be used. As our insights pertain to the impact that the magnitude of numbers has on basic arithmetic operations (subtraction and division), our results are broadly applicable to any difference- and ratio-based bias measures, and also do not depend on the model or dataset used for evaluation. Our experimental setup resembles the evaluation scenarios of many prior studies. This made it suitable for our analysis of bias measures used in prior research. However, the VoxCeleb datasets, which have now been retracted by the authors, present numerous ethical and privacy concerns~\cite{rusti2023benchmark}. We thus do not recommend them for evaluation.
\section{Conclusion}

This paper studied the impact of base metrics, bias measures and meta-measures on the outcomes of speaker verification bias evaluations. Our empirical analysis demonstrates that metrics and measures significantly impact evaluation outcomes. We recommend the use of ratio-based bias measures, in particular when the values of base metrics are small, or when base metrics with different orders of magnitude need to be compared. %We highlighted that the performance ranking across groups depends on the base metric used for evaluation. Ratio-based bias measures are more robust than difference-based measures, as they are not affected by the magnitude of the base metric. For interpretability we recommend bias measures that calculate a log ratio. Of the meta-measures we compared, the log ratio based NRB leads to a correct bias assessment, while the difference-based FDR underestimates bias.

\bibliographystyle{IEEEtran}
\bibliography{references}

\end{document}